\documentclass[
    ,final            
  ]
  {aipproc}

\layoutstyle{6x9}
 \newcommand {\nc} {\newcommand}
 \nc {\tstrut} {\vline  height 2.5ex depth 1.7ex width 0ex}
\begin{document}

\title{Polarisation Observables in
 Antiproton
\\
 Proton to Lepton Antilepton Reactions}



\author{Elise Jennings}{
  address={School of Mathematics, Trinity College Dublin, Ireland.}
}

\begin{abstract}
The electromagnetic form factors of hadrons as measured both in the space like and time like domains provide fundamental information on the nucleon structure and internal dynamics. General expressions, including the lepton mass, for the spin averaged differential cross section for the annihilation reaction lepton antilepton to proton antiproton are given, as well as general formulae for the single and double spin asymmetries. The time reversed reaction would involve a kinematic factor. We also present general expressions for the helicity amplitudes for this reaction. 
\end{abstract}

\maketitle

\section{Introduction}
 Electromagnetic form factors in the time like region are largely unknown and the relative phases have never been measured in the proton case.
These time like form factors are accessible through annihilation reactions such as $ \bar{p}\,p \to e^-\,e^+$ and its time reverse.
It is possible to measure the phases of the form factors via single spin asymmetry while measurements of the double  spin asymmetry will limit relative phase ambiguity and allow an independent $G_E$--$G_M$ separation.
\\
\noindent Recent measurements \cite{Jones:2000,Punjabi:2005wq} of the electron to proton polarisation transfer in $e^-\,p \to e^-\,p$ scattering at Jefferson Laboratory show that the ratio of Sachs form factors $G_E(q^2)/G_M(q^2)$ is monotonically decreasing with increasing $q^2$ in strong contradiction with the $G_E/G_M$ scaling assumed in the Rosenbluth separation method.
\\
\noindent It is thought that polarised antiprotons can explore this unexpected $q^2$
dependence of the ratio of electric and magnetic
form factors by studying their phases.
 In the case of mu and tau final state pairs it will be necessary to retain the lepton mass in the formulae for both the spin averaged differential cross section and the polarisation observables \cite{Buttimore:2006mq}.

\section{Spin averaged cross section}
The unpolarised differential cross section for $s$ channel annihilation of spin half particles in the centre of mass system is
\begin{eqnarray}
\frac{\sqrt{s - 4m_f^2}\,\,\sqrt{s - 4m_i^2}}{4\,\pi}\,\,\frac{d\sigma}{dt}&=&\frac{d\sigma}{d\Omega}\, = \, \frac{\beta}{64\pi^2 s}\,\frac{1}{4}\,\sum_{\mbox{\tiny spin}} |{\cal M}|^2
\end{eqnarray}
 where $\cal{M}$ is the invariant amplitude for the process and         $\beta$ is a flux factor. In an annihilation reaction of two spinor particles of mass $m_i$ producing a pair of mass $m_f$ the flux factor $\beta$ is given by
\vspace{-.75cm}
\begin{eqnarray}
 \beta& = &\left( \frac{s - 4\,m_f^2}{s - 4\,m_i^2}\right)^{1/2}.
\end{eqnarray}
%
%
%
 The spin averaged differential cross section for $l^+ + l^- \to p + \bar{p}\,$  scattering in terms of Mandelstam variables $s$ and $t$ is:
     \begin{eqnarray}
\label{unpol}
 \frac{d\sigma}{d\Omega}
 & =
 &\alpha^2 \,\beta\,\frac{1}{s^3\,\left(s - 4\,M^2\right)}\,
 \Bigg\{
 \,\frac{s^2}{2}\,\left(\,s - 4\,M^2 \right)
 \,|G_{M}|^2 \, -\,4\,s\,m^2\,M^2\,\left(\,|G_{M}|^2 - |G_{E}|^2\,\right)\nonumber \\[1ex]
     &&\qquad
+\,\left[\,\left(t\,-\,m^2 - M^2 \right)^2 + s\,t \right]\,\left(\,s\,|G_{M}|^2 - 4\,M^2|G_{E}|^2 \right)\,\Bigg\}
 \end{eqnarray}
 where $m$ is the mass of the lepton and $M$ is the mass of the         proton. The flux factor $\beta = \sqrt{s-4\,M^2}/\sqrt{s-4\,m^2}$ and $t = \left(P - K\right)^2$.
Dirac and Pauli form factors are normalised $F_{1}(0) = 1$ and $F_     {2}(0) = \mu_{p} - 1$, the anomalous magnetic moment and Sachs EM form factors are (with $\tau = s/4\,M^2$)
 \begin{eqnarray}
 G_{E}\,  =\,  F_{1} + \tau\,F_{2}\,, \qquad G_{M}\, =\,  F_{1} + F_{2}\,        .
 \end{eqnarray}
Neglecting the mass of the lepton in Eq. (\ref{unpol}) gives the cross section Ref.~\cite{Dubnickova:1992ii}.
    \begin{eqnarray}
  \frac{d\sigma}{d\Omega} & = &\frac{\alpha^2}{4}\,\frac{\beta}{s}\,\Bigg\{(1+\cos^{2} \theta)\,|G_{M}|^{2} + \, \frac{1}{\tau}\,\sin^{2}\theta\,|G_{E}|^{2}\Bigg\}
\\\nonumber
 \end{eqnarray}

\section{Asymmetries with lepton mass}
It is convenient to define a scaled unpolarised cross section, $D$, in terms of $s$ and $t$.
\begin{eqnarray}
\frac{d\sigma}{d\Omega}&=& \frac{\alpha^2}{4}\,\frac{\beta}{s}\,D\, 
\\[2ex]
D
 &=
 &\frac{16\,M^2}{s^2\,(s - 4\,M^2)}\,
 \bigg\{\,
\left[\,\left(\,t\,- \,m^2 - M^2 \right)^2 + s\,t \right]\,\left(\,\frac{s}{4\,M^2}\,|G_{M}|^2 - |G_{E}|^2 \right)\nonumber \\[2ex]
&&\qquad
 + \,\frac{s^2}{8\,M^2}\,\left(s - 4\,M^2\right)\,|G_{M}|^2 - s\,m^2\,\left(\,|G_{M}|^2 - |G_{E}|^2\right)\bigg\}\,.
\end{eqnarray}

\subsection{ Single Spin Asymmetry}
\noindent When the antiproton in $l^+\,l^- \to p\,\bar{p}$ is polarised and if the initial leptons are unpolarised we obtain the single spin asymmetry $A_{N}$ using the polarisation vector $S_{N}~=~(0,\,0,\,1,\,0)$.
 The asymmetry parameter $A_N$ is  defined as a measure of the         left-right asymmetry, which for ($m \not= 0$) and for either a polarised proton or antiproton is
\vspace{-.25cm}
\begin{eqnarray}
\label{A_N}
 A_{N}
 & = &
 \left(\,1 - \frac{4\,m^2}{s}\,\right)\,\frac{2\,M\,\sin 2\theta\,}{\sqrt{s}\,D}\,\mbox{Im}\,G_{E}^*G_{M}\,.
 \end{eqnarray}
 This is an example of how T-odd observables can be non zero if final state interactions give interfering amplitudes.
 \noindent For single photon exchange the other single spin observables $A_S$ and $A_L$ are non zero only when the initial lepton is polarised. The centre of mass scattering angle $\theta$ is given by
\vspace{-.25cm}
\begin{eqnarray}
\cos\theta &=& 
\frac{t - u}{\tstrut \sqrt{s - 4\,m^2}\,\sqrt{s - 4\,M^2}}\,.
\end{eqnarray}
\subsection{Double Spin Asymmetry}
\noindent
The double spin observables involving normal (N), transverse (S) and
longitudinal (L) spin directions are
 \begin{eqnarray}
 A_\mathrm{SS}
&=
 &\emph{N}\,\Bigg\{\left[\,\left(\,t\,- \,m^2 - M^2 \right)^2 + s\,t \right]\,\left(\,s\,|G_{M}|^2 - 4\,M^2|G_{E}|^2 \right)\,
\\[1ex]\nonumber
&&
 +\,\frac{1}{2}\,s\,(\,s - 4\,M^2\,)\, \left(\,s - 4\,m^2\right)\,\sin^2 \theta\,|G_M|^2 - 4\,s\,m^2\,M^2\, \left(\,|G_{M}|^2 - |G_{E}|^2\right)\Bigg\}
 \\[3ex]
 A_\mathrm{NN}
 & =
& \emph{N}\,\bigg\{\left[\,\left(\,t\,- \,m^2 - M^2 \right)^2 + s\,t \right]\,\left(\,s\,|G_{M}|^2 - 4\,M^2\,|G_{E}|^2 \right) \nonumber\\[1ex]
&&\qquad \qquad \qquad \qquad
\,- 4\,s\,m^2\,M^2\,\left(\,|G_{M}|^2 - |G_{E}|^2\right)
  \bigg\}
  \\[3ex]
  A_\mathrm{LL}
  & =
  & \emph{N}\,
  \bigg\{\left[\,\left(\,t\,- \,m^2 - M^2 \right)^2 + s\,t \right]\,\left(\,s\,|G_{M}|^2 + 4\,M^2\,|G_{E}|^2 \right)\,\nonumber
  \\[1ex]
  &&
 + \frac{1}{2}\,s^2\,(\,s - 4\,M^2\,)\, |G_{M}|^2 \, +\, 4\,s\,m^2\,M^2\,\left(\,|G_{M}|^2 - |G_{E}|^2\right)\bigg\}
\end{eqnarray}
where the coefficient $\emph{N}$ is given by
     \begin{eqnarray}
\emph{N}& =& \frac{ 4}{s^2\,\left(\,s - 4\,M^2\,\right)\,D}\,.
 \end{eqnarray}
 The last two double spin observables, $A_{SL}$ and $A_{LN}$, in terms of the centre of mass scattering angle are
     \begin{eqnarray}
 A_\mathrm{SL} & = & \frac{2\,M}{\sqrt{s}\,D}\,\left(\,1 - \frac{4\,m^2}{s}\,\right)\,\sin 2\theta\,\mbox{Re}\,G_{E}^*G_{M}
 \\[1ex]
 A_\mathrm{LN} & = & \frac{2\,M}{\sqrt{s}\,D}\,\left(\,1 - \frac{4\,m^2}{s}\,\right)\,\sin 2\theta\,\mbox{Im}\,G_{E}^{*}G_{M}\,.
\end{eqnarray}
\noindent
 Polarisation observables can be used to pin down the relative phases of the time like form factors.
 \noindent
 All of the double spin observables depend on the moduli squared of the form factors apart from $A_{SL}$ and $A_{LN}$ which contain the real and imaginary parts, respectively.
 All of the above formulae  reduce to previously published expressions \cite{Gakh:2005hh} for double spin observables when we neglect the mass of the lepton.
\section{Helicity Amplitudes}
Using  P, T and C invariance we find that there are five independent helicity amplitudes for $\bar{p}\,p \to l^-\,l^+$.
\vspace{-.1cm}
 \begin{eqnarray}
H\,(+\,+\,+\,+) &=&H\,(+\,+\,-\,-) = \phantom{-}H\,(-\,-\,-\,-) =  \phantom{-}H\,(-\,-\,+\,+)\nonumber
\\
H\,(+\,+\,-\,+) &=&  H\,(-\,-\,-\,+)= -\,H\,(+\,+\,+\,-) =-\,H\,(-\,-\,+\,-)\nonumber
 \\
H\,(+\,-\,-\,-) &=& H\,(+\,-\,+\,+) = -\,H\,(-\,+\,-\,-) =-\,H\,(-\,+\,+\,+)\nonumber
\\
H\,(+\,-\,+\,-) &=& H\,(-\,+ \,-\,+)\nonumber
\\
H\,(+\,-\,-\,+) &=& H\,(-\,+ \,+\,-)
\end{eqnarray}
where
\vspace{-.25cm}
 \begin{eqnarray}
 \frac{s}{\beta}\,H\,(+\,+\,+\,+) &=& - 2\,\frac{\alpha\,m\,M}{s}\,\cos\theta\,G_E\nonumber
 \\[1ex]
 \frac{s}{\beta}\, H\,(+\,+\,-\,+) &=&  \phantom{-}\frac{\alpha\,M}{\sqrt{s}}\,\sin\theta\, G_E\nonumber
  \\[1ex]
 \frac{s}{\beta}\,H\,(+\,-\,-\,-) &=& \phantom{-}
 \frac{\alpha\,m}{\sqrt{s}}\,\sin\theta\,G_M
 \\[1ex]
 \frac{s}{\beta}\,H\,(+\,-\,+\,-) &=&\nonumber
- \frac{\alpha}{2}\, \left( 1 + \cos\theta \right)\,\nonumber
  G_M\nonumber
   \\[1ex]
   \frac{s}{\beta}\,H\,(+\,-\,-\,+) &=&
 -  \frac{\alpha}{2}\,\left( 1 - \cos\theta \right)\,
   G_M\,.\nonumber
 \end{eqnarray}


\begin{theacknowledgments}
The author is grateful to the Irish Research Council for Science, Engineering and Technology
 (IRCSET) for a postgraduate research studentship and Trinity College Dublin for the award of a Scholarship.

\end{theacknowledgments}

\vspace{-1cm}
\bibliographystyle{aipproc}   

\begin{thebibliography}{9}


 \bibitem{Jones:2000}
  Jefferson Lab Hall A Collaboration, M.~K.~Jones \emph{et al.}, Phys. Rev. Lett. \textbf{84}, 1398 (2000); Jefferson Lab Hall A Collaboration, O.~Gayou \emph{et al.},\emph{ibid}. \textbf{88}, 092301 (2001)
  %
  \bibitem{Punjabi:2005wq}
  V.~Punjabi {\it et al.},
 Phys.\ Rev.\ C \textbf{ 71} (2005) 055202.
 [Erratum-ibid.\ C \textbf{ 71} (2005) 069902]
  [arXiv:nucl-ex/0501018].

\bibitem{Buttimore:2006mq}
  N.~H.~Buttimore and E.~Jennings,
  arXiv:hep-ph/0607227.


 \bibitem{Dubnickova:1992ii}
  A.~Z.~Dubnickova, S.~Dubnicka and M.~P.~Rekalo,
 Nuovo Cim.\ A \textbf{ 109} (1996) 241.



 \bibitem{Gakh:2005hh}
 G.~I.~Gakh and E.~Tomasi-Gustafsson,
 Nucl.\ Phys.\ A \textbf{ 771} (2006) 169.
 [arXiv:hep-ph/0511240].

\end{thebibliography}

 \end{document}